\documentclass[copyright,creativecommons]{eptcs}

\providecommand{\event}{ACL2-2022}

\usepackage{alltt}
\usepackage{enumitem}
\usepackage[T1]{fontenc}
\usepackage{inconsolata}
\usepackage{multirow}
\usepackage{underscore}
\usepackage{url}
\usepackage{xcolor}

\newcommand{\secref}[1]{Section \ref{#1}}

\newcommand{\tabref}[1]{Table \ref{#1}}

\newcommand{\citecode}[2]
{\cite[\href{https://github.com/acl2/acl2/tree/master/#1}{\texttt{#2}}]
    {acl2-code}}
\newcommand{\citeman}[2]
{\cite[\href{http://acl2.org/manual?topic=#1}{\texttt{#2}}]
    {acl2-manual}}

\newcommand{\code}[1]{\texttt{#1}} % in text
\newenvironment{bcode} % blocks
 {\begin{quote}\small\begin{alltt}}
 {\end{alltt}\end{quote}}
\definecolor{commentgray}{gray}{0.4}
\newcommand{\ccode}[1]{\texttt{\textcolor{commentgray}{#1}}} % comments

\begin{document}

%%%%%%%%%%%%%%%%%%%%%%%%%%%%%%%%%%%%%%%%%%%%%%%%%%%%%%%%%%%%%%%%%%%%%%%%%%%%%%%%

\title{A Complex Java Code Generator for ACL2 \\
       Based on a Shallow Embedding of ACL2 in Java}

\author{Alessandro Coglio
        \institute{Kestrel Institute \\ \url{http://www.kestrel.edu}}}

\def\titlerunning{Java Code Generation}
\def\authorrunning{A. Coglio}

\maketitle

\begin{abstract}
This paper describes a code generator
that translates ACL2 constructs to corresponding Java constructs,
according to a shallow embedding of ACL2 in Java.
Starting from purely functional ACL2 code,
the generated Java code exhibits imperative and object-oriented features
like destructive updates, loops, and overloading.
The overall translation from ACL2 to Java is fairly elaborate,
consisting of
several ACL2-to-ACL2 pre-translation steps,
an ACL2-to-Java proper translation step,
and several Java-to-Java post-translation steps.
Experiments suggest that the generated Java code
is not much slower than the ACL2 code.
The code generator can also recognize, and translate to Java,
ACL2 representations of certain Java constructs,
forerunning a code generation approach based on
a shallow embedding of Java in ACL2 (i.e.\ going the other way).
This code generator builds upon, and significantly extends,
a simple Java code generator for ACL2
based on a deep embedding of ACL2 in Java.
\end{abstract}

%%%%%%%%%%%%%%%%%%%%%%%%%%%%%%%%%%%%%%%%%%%%%%%%%%%%%%%%%%%%%%%%%%%%%%%%%%%%%%%%

\section{Introduction}
\label{intro}

The work reported in this paper builds upon, and significantly extends,
the work reported in the ACL2-2018 Workshop paper
``A Simple Java Code Generator Based on a Deep Embedding of ACL2 in Java''
\cite{atj-deep}.
That paper described a relatively simple (and somewhat unconventional) approach
to run ACL2 code in Java
by running an ACL2 interpreter written in Java
on a Java representation of the ACL2 code,
where the interpreter is a deep embedding of ACL2 in Java.
This paper describes a relatively complex (and more conventional) approach
to run ACL2 code in Java
by translating ACL2 code to suitably equivalent Java code
and running that Java code,
where the translation is a shallow embedding of ACL2 in Java.

Notwithstanding ACL2's inherent ability to run its code efficiently as Lisp,
motivations for generating code in other languages from ACL2 code
were provided in \cite{atj-deep}:
in some situations,
the code must interoperate with external code written in those languages
in an integrated and efficient way.
Moreover, certain applications may require
code written in specific languages,
such as C for embedded systems or device drivers.
In formal program synthesis by stepwise refinement
\cite{bmethod,zed,vdm,specware-www},
a derivation of an implementation from a high-level specification
typically ends with the translation of
the low-level (i.e.\ fully refined) specification
to a program in some language:
thus, code generators for ACL2 complement APT
\cite{apt-www,apt-simplify,apt-isodata},
for synthesizing code in languages other than ACL2.

The shallow embedding approach to Java code generation described in this paper
is realized in ATJ (\textbf{A}CL2 \textbf{T}o \textbf{J}ava)
\citeman{JAVA____ATJ}{java::atj}.
ATJ operates in two main modes,
corresponding to the deep and shallow embedding approaches.
The two modes share some features,
but the new mode provides a significantly novel capability.
The generated Java code is much more efficient and idiomatic
than the version reported in \cite{atj-deep}.
The translation from ACL2 to Java is fairly elaborate,
and involves some interesting renderings of ACL2 concepts as Java concepts.
Starting from purely functional ACL2 code,
the generated Java code exhibits imperative and object-oriented features
like destructive updates, loops, and overloading.
The new end-to-end translation consists of
several ACL2-to-ACL2 pre-translation steps,
an ACL2-to-Java proper translation step,
and several Java-to-Java post-translation steps;
some of the pre-translation and post-translation steps
may be of independent interest.
Besides the ability to translate ACL2 constructs to Java constructs
according to the shallow embedding of ACL2 in Java,
the new code generation mode provides
the ability to recognize, and translate to Java,
ACL2 representations of certain Java constructs,
according to a shallow embedding of Java in ACL2 (i.e.\ going the other way);
the latter ability foreruns the code generation approach
more fully realized in ATC \cite{atc} \citeman{C____ATC}{c::atc}.

ATJ has been and is being used in several applications at Kestrel,
such as safety monitors for learning-enabled components in avionics systems
\cite{lec-monitors}
and cryptographic functions.

\secref{features} describes the main features of the code generator.
\secref{performance} presents performance measurements on the generated code.
\secref{future} discusses future work.
\secref{related} surveys related work.

%%%%%%%%%%%%%%%%%%%%%%%%%%%%%%%%%%%%%%%%%%%%%%%%%%%%%%%%%%%%%%%%%%%%%%%%%%%%%%%%

\section{Features of the Code Generator}
\label{features}

Due to space limitations, this section only provides an overview of ATJ.
More detailed information is in the tutorial and reference documentation,
as well as in the documentation of the implementation.

%%%%%%%%%%%%%%%%%%%%%%%%%%%%%%%%%%%%%%%%

\subsection{Modes of Operation}
\label{modes}

ATJ operates in two main modes,
corresponding to the deep and shallow embedding;
each mode has a sub-mode
based on whether the satisfaction of guards is assumed or not.
There are thus four combinations:
(i) deep embedding not assuming guards;
(ii) deep embedding assuming guards;
(iii) shallow embedding not assuming guards; and
(iv) shallow embedding assuming guards.

The first combination was described in \cite{atj-deep},
and has not changed much since then.
The second combination is new, but not very different from the first one.
The third and fourth combinations are new;
the fourth one is far more interesting than the third one,
and is thus the main subject of this paper,
which nonetheless also briefly discusses the other three combinations.

Here `assuming guards' means that
the generated Java code mimics the ACL2 code
only when each ACL2 function is called
on arguments that satisfy the function's guard.
ATJ does not check this assumption,
and does not require the ACL2 code to be guard-verified;
in fact, ATJ also accepts program-mode ACL2 code.
If this guard assumption is violated,
ATJ does not guarantee the correctness of the generated Java code.%
\footnote{Since the generated Java code is not formally verified yet,
ATJ's guarantees are only as strong as
the author's best effort to design and implement ATJ correctly,
and the confidence derived from the current working ATJ tests.}
Running the generated Java code without knowing if the guard assumption holds
should be no more troubling than
running program-mode or non-guard-verified ACL2 code
without checking guards at run time
\citeman{ACL2____EVALUATION}{evaluation}.
If the ACL2 code is guard-verified,
and external Java code only calls
Java methods generated from ACL2 functions
whose guards are no stronger than the Java types of the methods' parameters,
then running the Java code is as good as running guard-verified ACL2 code.

%%%%%%%%%%%%%%%%%%%%%%%%%%%%%%%%%%%%%%%%

\subsection{Use of the Deep Embedding}
\label{deep}

AIJ (\textbf{A}CL2 \textbf{I}n \textbf{J}ava)
is the ACL2 interpreter written in Java
(i.e.\ the deep embedding of ACL2 in Java)
described in \cite{atj-deep}.
AIJ includes a Java representation of the ACL2 values
and a Java implementation of a collection of built-in ACL2 functions
that includes all the primitive ACL2 functions
\citeman{ACL2____PRIMITIVE}{primitive}
and that ATJ designates as `natively implemented ACL2 functions'.%
\footnote{Here `native' is from ACL2's point of view, not Java's.
These ACL2 functions are natively implemented in Java,
instead of being interpreted like the other ACL2 functions.
This is unrelated to Java native methods,
which are Java methods not written in Java;
AIJ is entirely written in Java.}
The AIJ interpreter executes
ACL2 terms and functions represented in Java
by moving around ACL2 values represented in Java
and by calling the natively implemented ACL2 functions.
The deep embedding mode of ATJ translates
the non-natively-implemented ACL2 functions
to Java code that constructs
Java representations of those functions' definitions
that can be executed by the interpreter.

The shallow embedding mode of ATJ uses a portion of AIJ:
not the representation and interpretation of the ACL2 terms and functions,
but the representation of the ACL2 values
and the natively implemented ACL2 functions.
This portion of AIJ is shared by the deep and shallow embedding.

This portion of AIJ was described, along with the rest of AIJ,
in \cite{atj-deep}.
AIJ has undergone numerous improvements and optimizations since then,
but that paper still provides a good overview.
More current and detailed information is in
the Javadoc documentation of the Java code of AIJ
\citeman{JAVA____AIJ}{java::aij}.

This shared portion of AIJ
provides Java classes \code{Acl2Integer}, \code{Acl2ConsPair}, etc.,
whose instances represent ACL2 values of the corresponding types.
These classes are hierarchically organized according to
the set containment relations among ACL2 values,
e.g.\ \code{Acl2Integer} is a subclass of \code{Acl2Rational},
and \code{Acl2Value} is a superclass of all the others.
These classes include methods to operate on these values,
e.g.\ to perform arithmetic on the numeric values;
these methods provide the core functionality
of the natively implemented ACL2 functions.

%%%%%%%%%%%%%%%%%%%%%%%%%%%%%%%%%%%%%%%%

\subsection{ACL2 Code Translated to Java}
\label{acl2-translated}

ATJ is given one or more target function symbols
that specify the ACL2 functions to be translated to Java.
ATJ translates not only those functions,
but also the ones transitively called,
stopping at the natively implemented ACL2 functions.

All the functions to be translated to Java must be defined,
i.e.\ have an unnormalized body
\citeman{ACL2____FUNCTION-DEFINEDNESS}{function-definedness},
or be constrained but have an attachment
\citeman{ACL2____DEFATTACH}{defattach};
ATJ treats the latter case
as if the constrained function were defined to call the attachment.
Furthermore, the functions must not have raw Lisp code
unless they are in a whitelist of functions known to be free of side effects.%
\footnote{A discussion of functions with raw Lisp code and side effects
is in \cite{atj-deep}.}
The built-in function \code{return-last} is handled specially
(see \secref{acl2-simplification}).
Lastly, the functions must not manipulate stobjs, as they involve side effects.
There are no other restrictions on the ACL2 functions translated to Java;
they may be in program or logic mode, guard-verified or not.

ATJ operates on the unnormalized bodies of the functions,
which are terms in translated form \citeman{ACL2____TERM}{term}.
Since untranslated terms may involve user-defined macros unknown to ATJ,
operating on translated terms is the appropriate approach for ATJ.
A downside is that ATJ needs to reconstruct some information
in the untranslated terms that gets lost in translation:
calls of \code{and} and \code{or},
which are easy to reconstruct from the calls of \code{if}
that they translate to;
and calls of \code{mv},
which require
an analysis of the number of results returned by ACL2 terms and functions,
and the recognition of the translated forms of \code{mv-let},
but are not too difficult to reconstruct this way.

%%%%%%%%%%%%%%%%%%%%%%%%%%%%%%%%%%%%%%%%

\subsection{Java Code Translated from ACL2}
\label{java-translated}

ATJ generates a main Java class,
and an auxiliary Java class that builds
a Java representation of the needed ACL2 environment.
ATJ also optionally generates a testing Java class
(see \secref{test-generation}).

In the deep embedding mode,
the auxiliary Java class builds a representation of
the ACL2 packages and functions.
The main Java class provides a wrapper
to call the interpreter on those functions.

In the shallow embedding mode,
the auxiliary Java class builds a representation of the ACL2 packages only.
The main Java class contains static methods
that correspond to the ACL2 functions.
These methods are organized into nested classes, one for each ACL2 package.
For instance, functions \code{f} and \code{g}
in package (i.e.\ whose \code{symbol-package-name} is) \code{"P"}
and \code{f} in package \code{"Q"}
are rendered as follows:
\begin{bcode}
public class ... \{ \ccode{// main class}
    public static class P \{ \ccode{// nested class for "P"}
        public static ... f(...) \{ ... \} \ccode{// method for f in "P"}
        public static ... g(...) \{ ... \} \ccode{// method for g in "P"}
        ... \ccode{// methods for other functions in "P"}
    \}
    public static class Q \{ \ccode{// nested class for "Q"}
        public static ... f(...) \{ ... \} \ccode{// method for f in "Q"}
        ... \ccode{// methods for other functions in "Q"}
    \}
    ... \ccode{// nested classes for other packages}
\}
\end{bcode}

This organization matches
the ACL2 package structure and
the use of package prefixes in ACL2.
In class \code{P}, methods \code{f} and \code{g}
can designate each other by their simple names,
just like the functions \code{f} and \code{g}
can be designated without package prefix within package \code{"P"}.
In class \code{Q}, method \code{f} must use
the qualified names \code{P.f} and \code{P.g}
to designate those methods in class \code{P},
just like the functions \code{p::f} and \code{p::g}
must be designated with that package prefix within package \code{"Q"}.

However, if package \code{"Q"} imports symbol \code{g} from package \code{"P"},
in ACL2 that symbol can be designated as just \code{g}, without package prefix,
within package \code{"Q"} as well.
The Java code generated by ATJ mimics this situation
by generating an additional method \code{g} in class \code{Q}
that acts as a ``synonym'' of \code{P.g} within class \code{Q},
by calling \code{P.g} and returning the result:%
\footnote{A JIT compiler in the JVM may inline calls of \code{Q.g}.}
\begin{bcode}
public class ... \{ \ccode{// main class}
    ... \ccode{// nested class P as above}
    public static class Q \{ \ccode{// nested class for "Q"}
        public static ... f(...) \{ ... \} \ccode{// method for f in "Q"}
        public static ... g(...) \{ return P.g(...); \} \ccode{// synonym of P.g}
        ... \ccode{// methods for other functions in "Q"}
    \}
    ... \ccode{// nested classes for other packages}
\}
\end{bcode}
This way, method \code{f} in class \code{Q} can use the simple name \code{g}
to designate (the synonym of) \code{P.g}.

ACL2 package, function, and variable names may use ASCII characters
that cannot be used in Java identifiers.
ATJ uses a fairly elaborate way to translate ACL2 names to Java names
that are valid identifiers, free of conflicts, and relatively idiomatic.
For instance, an ACL2 package name \code{"JAVA-VM"}
is translated to a Java class name \code{JAVA_VM},
while an ACL2 function or variable \code{square-root},
whose \code{symbol-name} is \code{"SQUARE-ROOT"},
is translated to a Java method or variable name \code{square_root}.%
\footnote{For these two examples, the names \code{JavaVM} and \code{squareRoot}
would be even more idiomatic in Java,
but they would require an even more elaborate name translation,
which is future work.}

ATJ represents the generated Java code
via an abstract syntax of (a sufficient subset of) Java,
defined via algebraic fixtypes \citeman{ACL2____FTY}{fty}.
The Java files are generated via a pretty-printer,
which minimizes parentheses in expressions
by considering the relative precedence of the Java expression constructs.

%%%%%%%%%%%%%%%%%%%%%%%%%%%%%%%%%%%%%%%%

\subsection{Architecture of the Translation}
\label{architecture}

The end-to-end translation from ACL2 to Java
is realized via a sequence of phases:
\begin{enumerate}[nosep]
\item
A pre-translation phase,
consisting of several ACL2-to-ACL2 transformation steps.
\item
A proper translation phase,
consisting of a single ACL2-to-Java transformation step.
\item
A post-translation phase,
consisting of several Java-to-Java transformation steps.
\end{enumerate}

This architecture reduces the overall complexity,
by decomposing it into more manageable pieces.
In particular, this simplifies the proper translation step,
which crosses the chasm between
two very different languages like ACL2 and Java:
the pre-translation steps move the ACL2 code towards Java,
and the post-translation steps move
the still somewhat ``ACL2-like'' Java code resulting from the translation
towards more efficient and idiomatic Java.

The pre-translation steps
simplify the ACL2 code according to certain criteria
(see \secref{acl2-simplification}),
reconstruct some information lost in ACL2's term translation
(see \secref{acl2-translated}),
annotate the ACL2 code
with type information
(see \secref{typed-translation} and \secref{overloaded-translation})
and variable reuse information
(see \secref{term-translation}),
and ensure that the ACL2 code satisfies certain conditions
(see \secref{inverse-shallow})
that are part of ATJ's input validation
but require a deeper analysis than possible
during ATJ's initial input processing.

The post-translation steps make the Java code more readable and efficient
(see \secref{java-simplification}).
In particular, they perform tail recursion elimination,
i.e.\ they turn tail recursions into loops.

All of the above mainly applies to the shallow embedding mode of ATJ.
In the deep embedding mode,
there are only a few pre-translation steps to simplify the code,
there are no post-translation steps,
and the proper translation step is fairly simple.

%%%%%%%%%%%%%%%%%%%%%%%%%%%%%%%%%%%%%%%%

\subsection{ACL2 Code Simplification}
\label{acl2-simplification}

A pre-translation step turns
\code{mbe} terms into their \code{:exec} or \code{:logic} parts,
and \code{mbt} terms into \code{t} or their arguments,
based on whether guards are assumed or not (see \secref{modes}).
This is the only difference between
the two sub-modes of the deep embedding mode.
In the guard-assuming sub-modes of
both deep embedding and shallow embedding modes,
this transformation can greatly increase performance.

When \code{mbt} terms are replaced with \code{t},
the resulting ACL2 code may include \code{if} calls with \code{t} tests,
which another pre-translation step turns into just their second arguments.

Since ATJ only accepts ACL2 code without side effects,
a pre-translation step turns \code{prog2\$} and \code{progn\$}
terms into their last arguments.

In translated ACL2 terms,
\code{mbe}, \code{mbt}, \code{prog2\$}, and \code{progn\$}
show up as \code{return-last} calls of certain forms.
These are the only forms of \code{return-last} currently accepted by ATJ.

A pre-translation step removes \code{let} bindings whose variables are not used.

Lambda expressions in ACL2 terms are always closed,
e.g.\ the term \code{(let ((x 0)) (cons x y))}
is \code{((lambda (x y) (cons x y)) '0 y)} in translated form,
where \code{y} is added, as both a parameter and an argument,
to the term \code{((lambda (x) (cons x y)) '0)},
which would be a more direct translation of the original \code{let} term
but would have an open lambda expression with free variable \code{y}.
This closure is beneficial for certain purposes,
e.g.\ it simplifies substitutions in translated terms
by making variable capture impossible.
However, for ATJ's purposes,
these extra variables like \code{y} add complications:
ATJ would treat the term with the closed lambda expression above
as \code{(let ((x '0) (y y)) (cons x y))},
but in doing so it would have to remove from consideration, on the fly,
the trivial binding of \code{y} to \code{y}
when translating the term to Java and analyzing variables for reuse
(see \secref{term-translation}).
Thus, a pre-translation step removes these trivial \code{let} bindings,
resulting in possibly open lambda expressions,
which are normal in functional languages,
and are in fact not present in ACL2 untranslated terms.
This makes the translation to Java and the variable reuse analysis simpler.

Some of the above transformations will need to be refined
when ATJ is extended to accept ACL2 code with side effects.
The guard satisfaction of \code{mbe} and \code{mbt} amounts to logical equality,
which is not the same as evaluation equality in the presence of side effects.
The non-last arguments of \code{prog2\$} and \code{progn\$},
and the unused \code{let} bindings,
may have side effects that should be preserved in Java.

%%%%%%%%%%%%%%%%%%%%%%%%%%%%%%%%%%%%%%%%

\subsection{Untyped Translation}
\label{untyped-translation}

Consider the following ACL2 function:
\begin{bcode}
(defun f (x y)
  (declare (xargs :guard (and (acl2-numberp x) (acl2-numberp y))))
  (* x (+ y 3)))
\end{bcode}

In the shallow embedding mode, not assuming guards,
the following Java code is generated:
\begin{bcode}
public static class ACL2 \{
    public static Acl2Value binary\_plus(Acl2Value x, Acl2Value y) ...
    public static Acl2Value binary\_star(Acl2Value x, Acl2Value y) ...
    public static Acl2Value f(Acl2Value x, Acl2Value y) \{
        return binary\_star(x, binary\_plus(y, \$N\_3)); \ccode{// (* x (+ y 3))}
    \}
\}
private static final Acl2Integer \$N\_3 = Acl2Integer.make(3);
\end{bcode}
Method \code{f} corresponds to function \code{f},
while methods \code{binary\_plus} and \code{binary\_star}
are wrappers for the native implementations of
the primitive functions \code{binary-+} and \code{binary-*}.
Field \code{\$N\_3} caches the \code{Acl2Integer} object
that represents the ACL2 integer 3
(the `\code{\$N}' in the name stands for `number').
Despite the guard of function \code{f},
method \code{f} and the other methods are essentially ``untyped'':
all their inputs and outputs have
the type \code{Acl2Value} of all the ACL2 values (see \secref{deep}).

In the shallow embedding mode, assuming guards,
the following Java code is generated:
\begin{bcode}
public static Acl2Number binary\_plus(Acl2Number x, Acl2Number y) ...
public static Acl2Number binary\_star(Acl2Number x, Acl2Number y) ...
public static Acl2Value f(Acl2Value x, Acl2Value y) \{
    return binary\_star((Acl2Number) x, binary\_plus((Acl2Number) y, \$N\_3));
\}
\end{bcode}
Method \code{f} is still untyped,
but now there are typed methods \code{binary\_plus} and \code{binary\_star}.
Method \code{f} includes casts from \code{Acl2Value} to \code{Acl2Number},
which looks worse than before on the face of it,
but the reason is just that ATJ has
(built-in) type information for \code{binary-+} and \code{binary-*}
but not for function \code{f};
this is easily remedied, as explained next.

%%%%%%%%%%%%%%%%%%%%%%%%%%%%%%%%%%%%%%%%

\subsection{Typed Translation}
\label{typed-translation}

Type information about function \code{f}
is communicated to ATJ via the following event:
\begin{bcode}
(atj-main-function-type f (:anumber :anumber) :anumber)
\end{bcode}
This claims that, under the guard, \code{f}
has two \code{acl-numberp} inputs and one \code{acl2-numberp} output.
Since \code{f} is in logic mode,
this claim is attempted to prove as a theorem,
which succeeds easily in this case;
the type information about \code{f}
is stored in a table \citeman{ACL2____TABLE}{table},
which ATJ consults when generating Java code.
It is not necessary for \code{f} to be guard-verified,
even though it is in this case.
If \code{f} were in program mode,
the theorem would be skipped,
but the type information would be still stored in the table;
this should be no more troubling
than assuming guard satisfaction in program-mode functions.

With this type information for function \code{f} available,
ATJ generates the following Java code:
\begin{bcode}
... \ccode{// typed binary\_plus and binary\_star as before}
public static Acl2Number f(Acl2Number x, Acl2Number y) \{
    return binary\_star(x, binary\_plus(y, \$N\_3)); \ccode{// no casts}
\}
\end{bcode}
Method \code{f} has stronger types, and the casts are avoided;
the code is more readable and efficient.

In general, the macro \code{atj-main-function-type}
specifies the input and output types of ACL2 functions,
where functions returning multiple values \citeman{ACL2____MV}{mv}
have multiple output types.
The types are designated by keywords,
such as \code{:anumber} for ACL2 numbers,
\code{:ainteger} for ACL2 integers,
\code{:acons} for ACL2 \code{cons} pairs,
\code{:avalue} for (all) ACL2 values,
and so on;
there are also \code{:j...} keywords for Java types,
discussed in \secref{inverse-shallow}.
In the macro name,
`\code{type}' is singular because
it refers to the (higher-order, i.e.\ arrow) type of the function,
and the significance of `\code{main}'
is explained in \secref{overloaded-translation}.
ATJ's built-in type information for \code{binary-+} and \code{binary-*},
mentioned in \secref{untyped-translation},
derives from predefined \code{atj-main-function-type} events,
for these and other built-in ACL2 functions.

A pre-translation step uses
the type information from the \code{atj-main-function-type} events
for the ACL2 functions translated to Java
(defaulting to \code{:avalue} when there is no better information),
to carry out a type analysis of the ACL2 terms
and to augment them with type annotations.
The type analysis calculates types for all the ACL2 terms,
ensuring compatibility with their surroundings;
variables are annotated with their types,
and all terms are wrapped with type conversions
that either are identities or bridge compatible types.
The type of a function's formal parameter
is obtained from the \code{atj-main-function-type} event
(or is \code{:avalue} if there is no such event).
The type of a term passed as argument to a function
must be identical or convertible to
the type of the function's corresponding parameter.
These are just two examples of this type analysis and annotation process,
which also handles
quoted constants,
\code{let} bindings,
\code{if} calls whose branch types are merged,
terms returning multiple values,
and additional aspects mentioned in
\secref{overloaded-translation} and \secref{inverse-shallow}.

For the example function \code{f} above,
in the absence of the \code{atj-main-function-type} event,
as in \secref{untyped-translation},
the application of this pre-translation step on the body of \code{f}
yields an annotated (translated) term of the following form
(where some package prefixes are omitted for readability):
\begin{bcode}
([AN>AV] (binary-* ([AV>AN] [AV]x)
                   ([AN>AN] (binary-+ ([AV>AN] [AV]y)
                                      ([AI>AN] '3)))))
\end{bcode}
The type annotations use short designations of types:
`\code{AV}' for \code{:avalue},
`\code{AN}' for \code{:anumber}, and
`\code{AI}' for \code{:ainteger}.
The variables \code{x} and \code{y} are annotated with their types,
by being renamed to a form that includes the type \code{:avalue} for both.
Since the quoted constant \code{'3} has type \code{:ainteger},
but is passed as argument to \code{binary-+}
whose corresponding parameter has type \code{:anumber},
the quoted constant is wrapped with \code{[AI>AN]},
which indicates a conversion from \code{:ainteger} to \code{:anumber}
(the `\code{>}' in the name conveys the direction).
This conversion is always possible,
because an ACL2 integer is also an ACL2 number.
Similarly, the variable \code{[AV]y} is wrapped with \code{[AV>AN]},
which indicates a conversion from \code{:avalue} to \code{:anumber}.
This conversion is always possible under the guard satisfaction assumption,
because \code{binary-*} requires \code{acl2-numberp},
even though not all ACL2 values are ACL2 numbers.
Since \code{binary-+} returns an ACL2 number
and \code{binary-*} takes an ACL2 number,
the \code{binary-+} call is wrapped with \code{[AN>AN]},
which indicates an identity conversion.
The wrapping of variable \code{[AV]x} is similar to \code{[AV]y}.
The outermost wrapping is based on
the output type \code{:avalue} of \code{f}
and the output type \code{:anumber} of \code{binary-*}.

The type conversions in the terms above are not real functions.
Together with the type annotations for variables,
they provide ``instructions'' to ATJ's proper translation step.
The variables' types tell the translation step
which types to use for the corresponding Java variables.
The type conversions tell the translation step
to generate additional code to realize some of the conversions.
Identity conversions like \code{[AN>AN]} require no additional code.
Conversions like \code{[AI>AN]} and \code{[AN>AV]} require no additional code
because \code{Acl2Integer} is a subclass of \code{Acl2Number}
and \code{Acl2Number} is a subclass of \code{Acl2Value}.
Conversions like \code{[AV>AN]} require a cast,
shown in the Java code in \secref{untyped-translation}.

In the presence of the \code{atj-main-function-type} event,
the pre-translation step applied to function \code{f}
yields an annotated term of the following form:
\begin{bcode}
([AN>AN] (binary-* ([AN>AN] [AN]x)
                   ([AN>AN] (binary-+ ([AN>AN] [AN]y)
                                      ([AI>AN] '3)))))
\end{bcode}
Since the variables have type \code{:anumber},
the Java code has no casts, as shown earlier.

Since ACL2 terms may return multiple values,
the conversions may involve multiple types
to the left and right of the `\code{>}'.
These multiple types are translated to generated Java classes
whose instances hold multiple values.
For example, if an ACL2 function returns an integer and a symbol,
its return type is a class \code{MV\_Acl2Integer\_Acl2Symbol};
only the needed \code{MV\_...} classes are generated.
ACL2 terms returning multiple values
are used in \code{mv-let} terms and thus immediately decomposed.
This is mimicked in the generated Java code:
an \code{mv-let} is translated to Java code that
assigns the \code{MV\_...} class instance to a variable
and then immediately assigns its fields to the bound variables.
Therefore, the \code{MV\_...} class instances are only briefly needed;
they are created once and re-used (according to a singleton pattern)
mimicking ACL2's efficient non-allocation of memory for multiple values.

ATJ's type apparatus affords flexible representations of ACL2 values in Java.
In particular, ATJ has a type \code{:aboolean} for ACL2 booleans,
which are symbols in ACL2, for which ATJ has a type \code{:asymbol}.
ATJ maps \code{:asymbol} to the Java class type \code{Acl2Symbol},
but maps \code{:aboolean} to the Java primitive type \code{boolean}.
This way, booleans can be used instead of \code{Acl2Symbol} instances
(with the appropriate \code{atj-main-function-type} events),
e.g.\ in conditional tests,
making the Java code more efficient and idiomatic.
When an ACL2 boolean is used as an ACL2 symbol,
ATJ infers a \code{[AB>AY]} conversion
(where `\code{AB}' designates \code{:aboolean}
and `\code{AY}` designates \code{:asymbol}),
which generates Java code to turn \code{true} or \code{false}
into the \code{Acl2Symbol} instances for \code{t} or \code{nil}.
Conversely, when an ACL2 symbol
that is supposedly a boolean under the guard satisfaction assumption
is used as a boolean,
the inferred conversion \code{[AY>AB]}
generates Java code to do the opposite conversion.

Similarly to booleans, ACL2 characters and strings
are mapped to the the Java types \code{char} and \code{String},
and converted to/from the \code{Acl2Character} and \code{Acl2String} types,
which are both subtypes of \code{Acl2Value},
as needed (these conversions amount to wrapping and unwrapping).
In the future, ACL2 integers could be similarly mapped
to the Java type \code{BigInteger}
and wrapped/unwrapped to/from \code{Acl2Integer} as needed.
It may be also possible to use the Java primitive integer types
\code{byte}, \code{short}, \code{int}, and \code{long}
for ACL2 integers in suitable ranges;
see \secref{future}.

%%%%%%%%%%%%%%%%%%%%%%%%%%%%%%%%%%%%%%%%

\subsection{Overloaded Translation}
\label{overloaded-translation}

The example function \code{f} above
maps ACL2 numbers to ACL2 numbers,
but it also maps ACL2 rationals to ACL2 rationals,
and ACL2 integers to ACL2 integers.
The two latter facts can be claimed,
and ensured by proving theorems since \code{f} is in logic mode,
via the following events:
\begin{bcode}
(atj-other-function-type f (:arational :arational) :arational)
(atj-other-function-type f (:ainteger :ainteger) :ainteger)
\end{bcode}
In the macro name, `\code{other}' is in relation to
`\code{main}' in the macro described in \secref{typed-translation}:
while that is the main type of \code{f}, implied by the guard,
these are auxiliary types of \code{f},
which bear no relation with the guard,
and must be narrower (i.e.\ more specific) than the main type.
The auxiliary types are stored in the same table as the main types,
and consulted by ATJ when generating Java code.

Since the built-in \code{binary-+} and \code{binary-*}
satisfy closure properties similarly to \code{f},
there are similar predefined \code{atj-other-function-type} events
for \code{binary-+} and \code{binary-*}.

For each auxiliary function type,
ATJ generates an overloaded method with more specific types
than the method generated for the main function type.
For \code{binary-+} and \code{binary-*},
ATJ generates more type-specific wrappers
of more type-specific native implementations.

The resulting Java code has the following form:
\begin{bcode}
public static Acl2Number binary\_plus(Acl2Number x, Acl2Number y) ...
public static Acl2Rational binary\_plus(Acl2Rational x, Acl2Rational y) ...
public static Acl2Integer binary\_plus(Acl2Integer x, Acl2Integer y) ...
public static Acl2Number binary\_star(Acl2Number x, Acl2Number y) ...
public static Acl2Rational binary\_star(Acl2Rational x, Acl2Rational y) ...
public static Acl2Integer binary\_star(Acl2Integer x, Acl2Integer y) ...
public static Acl2Number f(Acl2Number x, Acl2Number y) \{
    return binary\_star(x, binary\_plus(y, \$N\_3));
\}
public static Acl2Rational f(Acl2Rational x, Acl2Rational y) \{
    return binary\_star(x, binary\_plus(y, \$N\_3));
\}
public static Acl2Integer f(Acl2Integer x, Acl2Integer y) \{
    return binary\_star(x, binary\_plus(y, \$N\_3));
\}
\end{bcode}
The bodies of the three \code{f} methods look the same,
but they call different overloaded
\code{binary\_plus} and \code{binary\_star} methods.
Java resolves method overloading by selecting the most specific types.

This `more specific' relation among Java types
is captured via a partial order on ATJ's types,
based on their mappings to Java types.
The ability of ACL2 functions to have multiple (main and auxiliary) types
affects the ATJ pre-translation step described in \secref{typed-translation}:
for each function call,
the most specific type is chosen
that does not require a down-conversion of the inputs;
if there is no such type,
the main function type is used, with the appropriate down-conversions.
ATJ also ensures that the main and auxiliary input types of each function
are closed under greatest lower bounds,
to ensure that Java can always find the most specific types.%
\footnote{The example ACL2 function \code{f}
could have additional auxiliary types besides the ones shown above.
For instance, it could have
one with inputs \code{:ainteger} and \code{:arational},
and one with inputs \code{:arational} and \code{:ainteger}.
Unless there is a type with inputs \code{:ainteger} and \code{:ainteger},
which are the greatest lower bound of the other two just mentioned,
ATJ stops with an error, prompting the user to add the missing types,
because otherwise Java would be unable to select
the most specific method when given two \code{Acl2Integer} arguments.}

This ability to generate overloaded methods can improve efficiency.
Arithmetic on \code{Acl2Integer}
is generally more efficient than on \code{Acl2Rational}.
Computations on Java booleans or characters or strings
are generally more efficient than on
\code{Acl2Symbol} or \code{Acl2Character} or \code{Acl2String}.

%%%%%%%%%%%%%%%%%%%%%%%%%%%%%%%%%%%%%%%%

\subsection{Translation of Terms to Expressions and Statements}
\label{term-translation}

The body of the example function \code{f} above
is translated to a single Java expression returned by method(s) \code{f}.
More in general, the proper translation step turns each ACL2 term
into (i) a Java expression and (ii) zero or more Java statements,
where the statements must be executed before the expression.
The statements derive from ACL2 \code{let} bindings,
which are turned into Java local variable declarations and assignments,
and from ACL2 \code{if} calls,
which are turned into Java conditional statements.

Consider the following ACL2 function:
\begin{bcode}
(defun g (x y)
  (let ((z (cons x y)))
    (if (equal x y)
        x
      z)))
\end{bcode}
The proper translation step turns that into the following Java method:
\begin{bcode}
public static Acl2Value g(Acl2Value x, Acl2Value y) \{
    Acl2ConsPair z = cons(x, y); \ccode{// variable declaration}
    Acl2Value \$tmp1; \ccode{// declare IF variable}
    if (equal(x, y)) \{ \ccode{// check IF test}
        \$tmp1 = x; \ccode{// assign IF true result to IF variable}
    \} else \{
        \$tmp1 = z; \ccode{// assign If false result to IF variable}
    \}
    return \$tmp1; \ccode{// return IF result from IF variable}
\}
\end{bcode}
Variable \code{z} comes from the \code{let}.
Variable \code{\$tmp1} is internally generated:
it is the expression that the \code{if} term translates to,
along with the Java \code{if} statement,
whose branches assign (the Java representation of)
the value returned by the ACL2 \code{if} to \code{\$tmp1}.
If there were other \code{if} terms,
additional variables \code{\$tmp2}, \code{\$tmp3}, etc.\
would be internally generated, using a counter that is threaded through.

The above example suggests a general recursive translation pattern.
To translate a term to an expression and some statements,
first its subterms are translated to expressions and statements;
then the expressions for the subterms are combined
to generate the expression and statements for the term,
and the statements for the subterms and term are concatenated in order.
This is more complicated for \code{and} and \code{or} terms,
which are turned into \code{\&\&} and \code{||} expressions,
because any statements for the second conjunct or disjunct
must not be executed if the first conjunct or disjunct
suffices to determine the result:
the treatment is somewhat similar to \code{if} terms,
which are also non-strict.

The treatment of \code{let} variables is fairly elaborate.
A pre-translation step analyzes them to determine when they can be reused:
it marks each variable as either reused or not;
this is tricky, because scoping rules differ between ACL2 and Java.
This pre-translation step takes place
after the one described in \secref{typed-translation},
which annotates variables with their types,
because variables may be reused only if they have the same type.
The markings on the variables serve as ``instructions''
to the proper translation step:
if a variable is reused, a Java assignment is generated;
otherwise, a Java declaration is generated.

Consider the following ACL2 function:
\begin{bcode}
(defun h ()
  (let ((x 1))
    (let ((x (+ x 1)))
      (* 2 x))))
\end{bcode}
Since the first \code{x} is only used
in the term to which the second \code{x} is bound,
the variable can be reused.
The annotated body of this function has the following form,
where the \code{lambda} expressions have been turned into \code{let} terms
for readability:
\begin{bcode}
([AV>AV]
 (let (([AI]x ([AI>AI] '1)))
   ([AV>AV] (let (([AI]x ([AI>AI] (binary-+ ([AI>AI] [AI]x)
                                            ([AI>AI] '1)))))
              ([AI>AV] (binary-* ([AI>AI] '2)
                                 ([AI>AI] [AI]x)))))))
\end{bcode}
The pre-translation step for variable reuse produces the following body,
where again the \code{lambda} expressions have been turned into \code{let} terms
for readability:
\begin{bcode}
([AV>AV]
 (let (([N][AI]x ([AI>AI] '1)))
   ([AV>AV] (let (([O][AI]x ([AI>AI] (binary-+ ([AI>AI] [N][AI]x)
                                               ([AI>AI] '1)))))
              ([AI>AV] (binary-* ([AI>AI] '2)
                                 ([AI>AI] [O][AI]x)))))))
\end{bcode}
The \code{[N]} marking means that the variable is new (i.e.\ not reused);
the \code{[O]} marking means that the variable is old (i.e.\ reused).
Thus, the following Java method is generated:
\begin{bcode}
public static Acl2Value h() \{
    Acl2Integer x = \$N\_1; \ccode{// variable declaration}
    x = binary\_plus(x, \$N\_1); \ccode{// variable assignment}
    return binary\_star(\$N\_2, x);
\}
\end{bcode}

A subsequent pre-translation step renames apart
variables that happen to have the same name (i.e.\ symbol) in ACL2
but that must be different variables in Java,
due to overlapping scopes and inability to reuse
(e.g.\ because the variables have different types).
The same pre-translation step also translates ACL2 variable names
to Java variable names (see \secref{java-translated}).

%%%%%%%%%%%%%%%%%%%%%%%%%%%%%%%%%%%%%%%%

\subsection{Java Code Simplification}
\label{java-simplification}

The proper translation step is uniform and systematic
in order to decrease its complexity,
but the resulting Java code may be verbose.
The verbosity is reduced via post-translation steps,
which simplify the Java code to be more efficient and idiomatic.

For example, as mentioned in \secref{term-translation},
\code{if} terms engender additional local variables,
such as \code{\$tmp1} in method \code{g}.
This approach is uniform and compositional,
but method \code{g} could be simplified
by folding the \code{return} statements into the \code{if} branches
and avoiding variable \code{\$tmp1} altogether.
This is done by a post-translation step,
producing the following method
for function \code{g} in \secref{term-translation}:
\begin{bcode}
public static Acl2Value g(Acl2Value x, Acl2Value y) \{
    Acl2ConsPair z = cons(x, y);
    if (equal(x, y)) \{
        return x; \ccode{// return folded here}
    \} else \{
        return z; \ccode{// return folded here}
    \}
\} \ccode{// no \$tmp1 variable}
\end{bcode}

Some post-translation steps turn tail recursions into loops,
a well-known compiler optimization known as `tail recursion elimination'.
The main post-translation step that achieves this operates as follows:
if a method is tail-recursive
(i.e.\ it calls itself only as its dynamically last action),
then its body is surrounded by a \code{while} loop with test \code{true},
and each recursive call is replaced with a \code{continue}
preceded by a parallel assignment of the recursive call arguments
to the formal parameters of the method.%
\footnote{The loop is equivalent to the tail recursion.
The parallel assignment and the \code{continue}
are equivalent to performing the recursive call.
The loop is exited via \code{return} like the recursion,
so the \code{true} loop test does not make the loop infinite.}
The parallel assignment is realized
by sequentializing it according to a topological dependency order when possible,
or otherwise by using temporary variables.
Two subsequent post-translation steps improve the loop as follows:
(i) if the loop body is an \code{if} statement of certain forms,
the continuation test is lifted from the \code{if} to the \code{while},
replacing the \code{true} test, which is more natural;
(ii) any \code{continue} not dynamically followed by other code in the loop body
is eliminated, as it is useless.
For example, consider the following tail-recursive factorial function
where \code{r} is the accumulator:
\begin{bcode}
(defun fact-tail (n r)
  (declare (xargs :guard (and (natp n) (natp r))))
  (if (zp n)
      r
    (fact-tail (1- n) (* n r))))
\end{bcode}
The corresponding Java method is the following:
\begin{bcode}
public static Acl2Integer fact\_tail(Acl2Integer n, Acl2Integer r) \{
    while (!zp(n)) \{
        r = binary\_star(n, r);
        n = binary\_plus(\$N\_minus1, n);
    \}
    return r;
\}
\end{bcode}

A post-translation step caches and reuses the result of
a method with no arguments (generated from a nullary function)
whose body is a single \code{return}
with an expression that does not call any method.
Instead of recomputing the same expression at every call,
a private static final field is added
that is initialized with the expression,
and the method is transformed to return that field.

%%%%%%%%%%%%%%%%%%%%%%%%%%%%%%%%%%%%%%%%

\subsection{Inverse Shallow Embedding}
\label{inverse-shallow}

The preceding subsections describe ATJ features
to render ACL2 constructs as Java constructs,
according to a shallow embedding of ACL2 in Java.
This subsection briefly discusses, via an example, ATJ's ability
to recognize certain Java constructs shallowly embedded in ACL2,
where the embedding thus goes the other way,
and to translate them to those Java constructs.
This approach started in ATJ, but was more fully developed in ATC.

Consider the following ACL2 function and event:
\begin{bcode}
(defun i (x y)
  (declare (xargs :guard (and (int-valuep x) (int-valuep y))))
  (int-add (int-mul (int-value 2) x)
           (int-mul y y)))
(atj-main-function-type i (:jint :jint) :jint)
\end{bcode}
Function \code{int-valuep} represents Java \code{int} values;
function \code{int-value} represents Java \code{int} literals;
functions \code{int-add} and \code{int-mul}
represent the Java \code{int} addition and multiplication operations.
These are part of an ACL2 model of the Java primitive types and operations
that is known to ATJ, and used to generate the following Java method:
\begin{bcode}
public static int i(int x, int y) \{
    return 2 * x + y * y;
\}
\end{bcode}
The correspondence between the ACL2 code and the Java code is clear.
The \code{int} types of the method are derived from
the \code{atj-main-function-type} event,
where \code{:jint} represents the Java \code{int} type;
other possibilities (for different functions)
are \code{:jbyte}, \code{:jshort}, etc.
The ACL2 code above looks like ``Java written in ACL2'',
i.e.\ it is Java shallowly embedded in ACL2.
Since this is not idiomatic ACL2 code,
APT transformations may be used to produce it from more idiomatic ACL2 code;
some APT transformations tailored to ATJ
have been and are being developed at Kestrel.

Similarly, an ACL2 model of Java primitive arrays and operations on them
is used to generate Java code that manipulates primitive arrays.
The model includes functions like
\code{int-arrayp} to recognize arrays of \code{int} values,
\code{int-array-read} to read from them, and
\code{int-array-write} to write to them.
A pre-translation step performs an analysis on the Java arrays
to ensure that they are treated in a single-threaded way
so that they can be destructively updated in the generated Java code.
The analysis is similar to ACL2's stobjs,
but involves additional complications:
array names are local to functions/methods,
and must be suitably mapped across calls,
while stobj names are global;
methods can create new arrays,
while stobjs are statically defined.
To specify how arrays are modified by functions/methods,
and to distinguish them from arrays created afresh,
array types used in \code{atj-main-function-type} outputs,
such as \code{:jbyte[]},
are augmented with names of array parameters
indicating which output array is which possibly modified input array.

This ability to generate Java code
that manipulates Java data types represented in ACL2
is integrated with the ability to generate Java code
that manipulates Java representations of ACL2 data types.
ATJ recognizes ACL2 conversion functions
between the models of the Java data types
and the ``regular'' ACL2 data types,
and generates appropriate Java conversion code.

A somewhat experimental option recently added to ATJ
triggers a new pre-translation step to ensure
that the ACL2 code only uses the models of the Java data types,
so that the generated Java code only traffics
in Java primitive values and arrays
(i.e.\ not \code{Acl2Integer} etc.).
This obviates the need for generating
the auxiliary class described in \secref{java-translated},
because if the Java code does not use \code{Acl2Symbol},
there is no need to create Java definitions of ACL2 packages;
it also obviates the need for generating some code in the main class,
such as fields like \code{\$N\_3} shown earlier.
With this option, ATJ works much more like ATC:
this option should evolve into a third main operation mode of ATJ,
based on the shallow embedding of Java in ACL2,
next to the modes based on the shallow and deep embedding of ACL2 in Java.
However, so far ATJ is still primarily based on the latter two modes.

%%%%%%%%%%%%%%%%%%%%%%%%%%%%%%%%%%%%%%%%

\subsection{Test Generation}
\label{test-generation}

In both the deep and shallow embedding modes,
ATJ provides a facility to generate additional Java code
to run tests on the Java code generated from the target functions.

Each test is specified by a name and
a ground call of one of the target functions supplied to ATJ.
For the example function \code{f} in \secref{untyped-translation},
an example test ground call is \code{(f 2 1)}.

If tests are specified,
ATJ generates an additional test class to run the tests.
For each test,
ATJ uses the ACL2 evaluator to calculate the result of the call
(e.g.\ \code{8} for the example above),
and generates Java code, in the test class, to
(i) build the input values,
(ii) call the Java counterpart of the ACL2 function on them,
(iii) compare the result with the one from the ACL2 evaluator, and
(iv) print whether the test passes or fails on the screen,
along with the name of the test.
Note that the result of the test does not have to be specified by the user.

This test class has a \code{main} method,
so it can be run as a Java application
to execute all the tests in the manner explained above.
If a positive integer argument is passed to this Java application,
the test code runs each test for the specified number of times,
collecting the execution times for each run
and printing maximum, minimum, and average times on the screen.
If a second positive integer argument is passed to this Java application,
it specifies the number of times to run the ground call
for each time measurement;
this is useful when the calls are fast,
to measure larger and less noisy times.

Since ATJ does not generate correctness proofs yet,
this testing facility is valuable to increase confidence in the code.
It is also useful to measure the performance of the generated code.

%%%%%%%%%%%%%%%%%%%%%%%%%%%%%%%%%%%%%%%%%%%%%%%%%%%%%%%%%%%%%%%%%%%%%%%%%%%%%%%%

\section{Performance of the Generated Code}
\label{performance}

Performance has been tested on
a verified ABNF grammar parser \cite{abnf} \citeman{ABNF____ABNF}{abnf::abnf}.
The parser takes as input a list of natural numbers,
checks whether they are ASCII codes
that form a grammar in the ABNF notation,
and returns as output a concrete syntax tree for the parsed grammar
(if parsing succeeds; otherwise it returns \code{nil}).
The grammars used in the tests either are taken from Internet standards
or are transcriptions in ABNF of existing grammars written in other notations.

The results are reported in \tabref{abnf-times}.
The first column describes the input grammars.
The other three columns show minimum, maximum, and average running times
for the ACL2 code of the parser,
with the default guard checking setting,%
\footnote{Namely, that
only the guard of the top-level call of the parser is checked,
and none of the guards of the directly or indirectly called functions
are checked.
The parser is all guard-verified.}
and for the Java code generated by ATJ
in the shallow and deep embedding modes, both assuming guards.

\begin{table}[ht]
\centering
{\footnotesize
\begin{tabular}{|r||rrr|rrr|rrr|}
\hline
\multicolumn{1}{|c||}{\multirow{2}{*}{\textbf{Input}}} &
\multicolumn{3}{c|}{\textbf{ACL2}} &
\multicolumn{3}{c|}{\textbf{Java\ \ \ [shallow]}} &
\multicolumn{3}{c|}{\textbf{Java\ \ \ [deep]}} \\
& \multicolumn{1}{c}{\textbf{min}}
& \multicolumn{1}{c}{\textbf{avg}}
& \multicolumn{1}{c|}{\textbf{max}}
& \multicolumn{1}{c}{\textbf{min}}
& \multicolumn{1}{c}{\textbf{avg}}
& \multicolumn{1}{c|}{\textbf{max}}
& \multicolumn{1}{c}{\textbf{min}}
& \multicolumn{1}{c}{\textbf{avg}}
& \multicolumn{1}{c|}{\textbf{max}} \\
\hline\hline
ABNF core rules &
0.030 & 0.033 & 0.047 &  % ACL2
0.078 & 0.127 & 0.261 &  % Java shallow
7.145 & 7.522 & 8.405 \\ % Java deep
\hline
ABNF grammar &
0.031 & 0.033 & 0.035 &  % ACL2
0.111 & 0.120 & 0.143 &  % Java shallow
7.540 & 7.704 & 8.132 \\ % Java deep
\hline
HTTP grammar &
 0.048 &  0.054 &  0.071 &  % ACL2
 0.188 &  0.200 &  0.250 &  % Java shallow
11.743 & 11.981 & 12.435 \\ % Java deep
\hline
IMAP grammar &
 0.292 &  0.337 &  0.364 &  % ACL2
 0.888 &  1.032 &  1.223 &  % Java shallow
71.384 & 72.103 & 72.837 \\ % Java deep
\hline
IMF grammar &
 0.122 &  0.137 &  0.159 &  % ACL2
 0.465 &  0.492 &  0.589 &  % Java shallow
29.961 & 30.466 & 30.971 \\ % Java deep
\hline
Java lexical grammar &
 0.158 &  0.172 &  0.189 &  % ACL2
 0.497 &  0.540 &  0.594 &  % Java shallow
36.849 & 37.238 & 40.028 \\ % Java deep
\hline
Java syntactic grammar &
  0.490 &   0.509 &   0.550 &  % ACL2
  1.546 &   1.654 &   1.863 &  % Java shallow
113.858 & 116.990 & 120.384 \\ % Java deep
\hline
JSON grammar &
0.027 & 0.032 & 0.048 &  % ACL2
0.114 & 0.124 & 0.147 &  % Java shallow
7.094 & 7.248 & 7.655 \\ % Java deep
\hline
SMTP grammar &
 0.128 &  0.143 &  0.165 &  % ACL2
 0.418 &  0.436 &  0.466 &  % Java shallow
30.259 & 30.571 & 31.264 \\ % Java deep
\hline
URI grammar &
 0.040 &  0.044 &  0.063 &  % ACL2
 0.155 &  0.163 &  0.174 &  % Java shallow
10.236 & 10.518 & 11.003 \\ % Java deep
\hline
Yul grammar &
 0.123 &  0.138 &  0.156 &  % ACL2
 0.420 &  0.438 &  0.465 &  % Java shallow
32.925 & 34.664 & 37.049 \\ % Java deep
\hline
\end{tabular}
}
\caption{\label{abnf-times}
Time measurements for the ABNF parser.}
\end{table}

For the Java code generated in shallow embedding mode,
a handful of \code{atj-main-function-type} events are used,
with more specific \code{:a...} types than \code{:avalue},
for certain lower-level functions in the parser.
All the other functions in the parser
have the default \code{:avalue} type for inputs and outputs;
no other (more specific) type can be used for them,
among the types currently supported by ATJ.
In other words, the Java types used in the Java code of the parser
are as specific as they can currently be.
The ACL2 code of the parser does not use
the models of Java primitive values or arrays
(see \secref{inverse-shallow}),
i.e.\ there are no \code{:j...} types
in the \code{atj-main-function-type} events;
the inverse shallow embedding plays no role here.
The details are in the Community Books
\citecode{books/kestrel/java/atj/tests/abnf.lisp}
         {books/kestrel/java/atj/tests/abnf.lisp}.

Each time in the table refers to
100 repeated calls of the parser on the same input,
to have larger and less noisy time measurements.
Minimum, maximum, and average times are calculated over 10 runs
(each run consisting of the 100 calls just explained).

Looking at the average times,
the Java code generated in shallow embedding mode
is about 3--4 times slower than the ACL2 code.
This is not insignificant,
but suggests that the Java code is somewhat competitive with the ACL2 code.
On the other hand,
the Java code generated in deep embedding mode
is about 60--80 times slower than
the Java code generated in shallow embedding mode.
This supports the claim that the shallow embedding mode
produces much more efficient Java code than the deep embedding mode.
It also suggests that the Java code generated in deep embedding mode
is not particularly competitive with the ACL2 code,
or with the Java code generated in shallow embedding mode.

It is not surprising that
the Java code generated in shallow embedding mode
is slower than the ACL2 code,
because the Java code is engineered to mimic the ACL2 code,
but the ACL2 code can take advantage of certain efficiencies
that are not available to the Java code.
A major instance of this is arithmetic:
ACL2 has unbounded integers,
which the underlying Lisp operates on very efficiently,
possibly via one machine instruction per operation
when they are sufficiently small;
the Java code mimics ACL2 unbounded integers
via the \code{BigInteger} class,
which is much less efficient than one machine instruction per operation,
no matter how small the integers.
In Java, arithmetic efficiency comparable to Lisp
can be achieved via primitive types like \code{int}:
currently, these can be generated by ATJ
only using the ACL2 models of the Java primitive values,
which requires transforming the parser's ACL2 code.
Future versions of ATJ may be able to generate Java code with primitive types
from ACL2 integers that are known to be sufficiently small
(see \secref{future});
this may reduce the performance gap with the ACL2 code.

The large disparity between deep and shallow embedding modes
is due to the interpretation overhead in the former compared to the latter.
A variable access in the interpreter is an array access
(variables are stored in arrays, and internally identified by indices),
compared to just referencing a variable.
Each function call in the interpreter first checks whether the function
is ACL2's \code{if} or \code{or},%
\footnote{Although \code{or} is a macro
that expands \code{(or a b)} to \code{(if a a b)},
AIJ represents it as a function internally
to evaluate \code{a} once.}
which require non-strict treatment,
compared to just calling a method.
Each function call in the interpreter
creates an array of the argument values,
compared to just passing them to the method.
These and other overheads add up.

All the time measurements were taken on
a MacBook Pro (16-inch, 2019)
with 2.4 GHz 8-Core Intel Core i9
and 64 GB 2667 MHz DDR4,
running macOS Big Sur Version 10.13.6.
The ACL2 times were measured with
GitHub commit e9e633f669ab49170c47538e400c6d30158ce5fb
running on Steel Bank Common Lisp (SBCL) Version 2.1.6.
The Java times were measured with the Java code generated by
the version of ATJ in the same GitHub commit,
running on OpenJDK 15 2020-09-15 (build 15+36).

%%%%%%%%%%%%%%%%%%%%%%%%%%%%%%%%%%%%%%%%%%%%%%%%%%%%%%%%%%%%%%%%%%%%%%%%%%%%%%%%

\section{Future Work}
\label{future}

Ideally, ATJ would automatically translate
idiomatic and efficient ACL2 code to idiomatic and efficient Java code.
While this may be impossible to achieve in full
due to the inherent differences between the two languages,
it is a good goal to move towards;
from ATJ's current status,
some movement in that direction should be achievable.
An approach worth exploring is the introduction
of additional \code{:a...} types
for ACL2 values that could be mapped to Java primitive and array types.
ATJ already does that for booleans, characters, and strings;
if this could be done for suitable subsets of the integers,
arithmetic could be much more efficient
(cf.\ discussion in \secref{performance}).
This applies to the shallow embedding of ACL2 in Java;
it is independent from ATJ's capability
to translate ACL2 models of Java primitive values and arrays to idiomatic Java,
which requires the ACL2 code to be in a specific form.

The latter capability
should be further developed into a new mode of operation of ATJ,
instead of being an add-on to the current shallow embedding mode,
which is primarily based on the shallow embedding of ACL2 in Java.
This new mode could work very similarly to C code generation in ATC,
according to a more comprehensive shallow embedding of Java in ACL2.
Both shallow embedding modes are useful, in different ways:
one puts fewer restrictions on the ACL2 code,
at the cost of producing less idiomatic and efficient Java code;
the other can produce more idiomatic and efficient Java code,
at the cost of having to transform the ACL2 code into a restricted form.

ATJ should be extended to accept ACL2 code with side effects
and translate it to Java code that mimics those side effects.
Since ATJ does not have access to
the raw Lisp code that realizes the side effects,
support for side effects can be added case by case,
by extending AIJ (see \secref{deep})
with Java implementations of ACL2 functions with side effects.
The side effects related to stobjs
could be handled in a more uniform way,
adding general support for stobjs.

ATJ should be extended to generate proofs of correctness
of the generated Java code with respect to the ACL2 code.
Each of ATJ's
pre-translation steps,
proper translation step, and
post-translation steps
could generate a proof of its own correctness,
obtaining an end-to-end proof by composing the proofs for all the steps.
ATJ translates to Java not only logic-mode ACL2 functions,
but also program-mode ones,
and possibly also ones with side effects in the future.
In order to assert, in the ACL2 logic,
the correctness of the generated Java code
with respect to ACL2 code in program mode and/or with side effects,
it is necessary to use a formal evaluation semantics of ACL2
\citeman{ACL2PL____ACL2-PROGRAMMING-LANGUAGE}
        {acl2pl::acl2-programming-language}.
 
Although the Java code generated in deep embedding mode is slow,
partial evaluation may make it more efficient.
According to the first Futamura projection \cite{parteval},
partially evaluating an interpreter with respect to a program
amounts to compiling the program
to the language that the interpreter is written in.
With a partial evaluator for Java,
AIJ could be partially evaluated with respect to
the ATJ-generated Java representation of the ACL2 code.
The partial evaluator for Java could be written in any language,
including ACL2:
in that case, it could be part of a Java transformation library
that ATJ's post-translation steps are a start towards;
the partial evaluator could generate proofs of correctness.

%%%%%%%%%%%%%%%%%%%%%%%%%%%%%%%%%%%%%%%%%%%%%%%%%%%%%%%%%%%%%%%%%%%%%%%%%%%%%%%%

\section{Related Work}
\label{related}

ATC \cite{atc} \citeman{C____ATC}{c::atc}
is a C code generator for ACL2
entirely based on the concept of inverse shallow embedding
that ATJ has pioneered but supports only partially.

Besides ATJ and ATC,
the author is not aware of any other code generator for ACL2.

Several other theorem provers include code generation facilities
\cite{coq-refman,isa-codegen,pvs-codegen}.
These are also based on direct shallow embedding,
but the ACL2 language is quite different from those provers' languages,
which are higher-order and strongly typed.
Thus, not many of the ideas from those provers' code generation facilities
may be relevant to ATJ.

Several formalizations of Java exist \cite{java-light,jbook,k-java}.
These may contain ideas relevant to extending
the partial model of Java mentioned in \secref{inverse-shallow}
to a formalization of Java in ACL2
sufficient to support formal proofs of correctness
of the generated Java code with respect to the ACL2 code,
which is part of the future work envisioned in \secref{future}.

%%%%%%%%%%%%%%%%%%%%%%%%%%%%%%%%%%%%%%%%%%%%%%%%%%%%%%%%%%%%%%%%%%%%%%%%%%%%%%%%

\section*{Acknowledgements}

Thanks to
Limei Gilham,
David Hardin,
Christoph Kreitz,
Eric McCarthy,
Karthik Nukala,
Eric Smith, and
Stephen Westfold
for using the Java code generator
and for providing valuable suggestions that led to improvements to the tool.
Thanks to
Eric Smith,
Karthik Nukala, and
Stephen Westfold
for developing new APT transformations
tailored to the inverse shallow embedding feature.

%%%%%%%%%%%%%%%%%%%%%%%%%%%%%%%%%%%%%%%%%%%%%%%%%%%%%%%%%%%%%%%%%%%%%%%%%%%%%%%%

\bibliographystyle{eptcs}
\bibliography{paper}

\begin{thebibliography}{10}
\providecommand{\bibitemdeclare}[2]{}
\providecommand{\surnamestart}{}
\providecommand{\surnameend}{}
\providecommand{\urlprefix}{Available at }
\providecommand{\url}[1]{\texttt{#1}}
\providecommand{\href}[2]{\texttt{#2}}
\providecommand{\urlalt}[2]{\href{#1}{#2}}
\providecommand{\doi}[1]{doi:\urlalt{http://dx.doi.org/#1}{#1}}
\providecommand{\eprint}[1]{arXiv:\urlalt{https://arxiv.org/abs/#1}{#1}}
\providecommand{\bibinfo}[2]{#2}

\bibitemdeclare{book}{bmethod}
\bibitem{bmethod}
\bibinfo{author}{Jean-Raymond \surnamestart Abrial\surnameend}
  (\bibinfo{year}{1996}): \emph{\bibinfo{title}{The {B-Book}: Assigning
  Programs to Meanings}}.
\newblock \bibinfo{publisher}{Cambridge University Press},
  \doi{10.1017/CBO9780511624162}.

\bibitemdeclare{inproceedings}{k-java}
\bibitem{k-java}
\bibinfo{author}{Denis \surnamestart Bogd{\u a}na{\c s}\surnameend} \&
  \bibinfo{author}{Grigore \surnamestart Ro{\c s}u\surnameend}
  (\bibinfo{year}{2015}): \emph{\bibinfo{title}{{K-Java}: A Complete Semantics
  of {Java}}}.
\newblock In: {\sl \bibinfo{booktitle}{Proc.\ 42nd {ACM} Symposium on
  Principles of Programming Languages ({POPL})}}, pp.
  \bibinfo{pages}{445--456}, \doi{10.1145/2676726.2676982}.

\bibitemdeclare{inproceedings}{lec-monitors}
\bibitem{lec-monitors}
\bibinfo{author}{Darren \surnamestart Cofer\surnameend}, \bibinfo{author}{Isaac
  \surnamestart Amundson\surnameend}, \bibinfo{author}{Ramachandra
  \surnamestart Sattigeri\surnameend}, \bibinfo{author}{Arjun \surnamestart
  Passi\surnameend}, \bibinfo{author}{Christopher \surnamestart
  Boggs\surnameend}, \bibinfo{author}{Eric \surnamestart Smith\surnameend},
  \bibinfo{author}{Limei \surnamestart Gilham\surnameend},
  \bibinfo{author}{Taejoon \surnamestart Byun\surnameend} \&
  \bibinfo{author}{Sanjai \surnamestart Rayadurgam\surnameend}
  (\bibinfo{year}{2020}): \emph{\bibinfo{title}{Run-Time Assurance for
  Learning-Based Aircraft Taxiing}}.
\newblock In: {\sl \bibinfo{booktitle}{Proc.\ 39th Digital Avionics Systems
  Conference ({DASC})}}, \doi{10.1109/DASC50938.2020.9256581}.

\bibitemdeclare{inproceedings}{abnf}
\bibitem{abnf}
\bibinfo{author}{Alessandro \surnamestart Coglio\surnameend}
  (\bibinfo{year}{2018}): \emph{\bibinfo{title}{A Formalization of the {ABNF}
  Notation and a Verified Parser of {ABNF} Grammars}}.
\newblock In: {\sl \bibinfo{booktitle}{Proc.\ 10th Working Conference on
  Verified Software: Theories, Tools, and Experiments ({VSTTE})}}, {\sl
  \bibinfo{series}{Lecture Notes in Computer Science (LNCS)}}
  \bibinfo{volume}{11294}, pp. \bibinfo{pages}{177--195},
  \doi{10.1007/978-3-030-03592-1_10}.

\bibitemdeclare{inproceedings}{atj-deep}
\bibitem{atj-deep}
\bibinfo{author}{Alessandro \surnamestart Coglio\surnameend}
  (\bibinfo{year}{2018}): \emph{\bibinfo{title}{A Simple Java Code Generator
  for {ACL2} Based on a Deep Embedding of {ACL2} in {Java}}}.
\newblock In: {\sl \bibinfo{booktitle}{Proc.\ 15th International Workshop on
  the {ACL2} Theorem Prover and Its Applications ({ACL2-2018})}}, {\sl
  \bibinfo{series}{Electronic Proceedings in Theoretical Computer Science
  (EPTCS)}} \bibinfo{volume}{280}, pp. \bibinfo{pages}{1--17},
  \doi{10.4204/EPTCS.280.1}.

\bibitemdeclare{inproceedings}{atc}
\bibitem{atc}
\bibinfo{author}{Alessandro \surnamestart Coglio\surnameend}
  (\bibinfo{year}{2022}): \emph{\bibinfo{title}{A Proof-Generating {C} Code
  Generator for {ACL2} Based on a Shallow Embedding of {C} in {ACL2}}}.
\newblock In: {\sl \bibinfo{booktitle}{Proc.\ 17th International Workshop on
  the {ACL2} Theorem Prover and Its Applications ({ACL2-2022})}}.

\bibitemdeclare{inproceedings}{apt-simplify}
\bibitem{apt-simplify}
\bibinfo{author}{Alessandro \surnamestart Coglio\surnameend},
  \bibinfo{author}{Matt \surnamestart Kaufmann\surnameend} \&
  \bibinfo{author}{Eric \surnamestart Smith\surnameend} (\bibinfo{year}{2017}):
  \emph{\bibinfo{title}{A Versatile, Sound Tool for Simplifying Definitions}}.
\newblock In: {\sl \bibinfo{booktitle}{Proc.\ 14th International Workshop on
  the {ACL2} Theorem Prover and Its Applications ({ACL2-2017})}}, {\sl
  \bibinfo{series}{Electronic Proceedings in Theoretical Computer Science
  (EPTCS)}} \bibinfo{volume}{259}, pp. \bibinfo{pages}{61--77},
  \doi{10.4204/EPTCS.249.5}.

\bibitemdeclare{inproceedings}{apt-isodata}
\bibitem{apt-isodata}
\bibinfo{author}{Alessandro \surnamestart Coglio\surnameend} \&
  \bibinfo{author}{Stephen \surnamestart Westfold\surnameend}
  (\bibinfo{year}{2020}): \emph{\bibinfo{title}{Isomorphic Data Type
  Transformations}}.
\newblock In: {\sl \bibinfo{booktitle}{Proc.\ 16th International Workshop on
  the {ACL2} Theorem Prover and Its Applications ({ACL2-2020})}}, {\sl
  \bibinfo{series}{Electronic Proceedings in Theoretical Computer Science
  (EPTCS)}} \bibinfo{volume}{327}, pp. \bibinfo{pages}{125--141},
  \doi{10.4204/EPTCS.280.1}.

\bibitemdeclare{misc}{coq-refman}
\bibitem{coq-refman}
\emph{\bibinfo{title}{Coq 8.15.0 Reference Manual}}.
\newblock \bibinfo{howpublished}{\url{https://coq.inria.fr}}.

\bibitemdeclare{misc}{isa-codegen}
\bibitem{isa-codegen}
\bibinfo{author}{\surnamestart {Florian Haftmann with contributions from Lukas
  Bulwahn}\surnameend} (\bibinfo{year}{2021}): \emph{\bibinfo{title}{Code
  generation from Isabelle/HOL theories}}.
\newblock \bibinfo{howpublished}{\url{https://isabelle.in.tum.de}}.
\newblock \bibinfo{note}{Tutorial distributed with Isabelle/HOL}.

\bibitemdeclare{book}{vdm}
\bibitem{vdm}
\bibinfo{author}{Cliff \surnamestart Jones\surnameend} (\bibinfo{year}{1990}):
  \emph{\bibinfo{title}{Systematic Software Development using {VDM}}},
  \bibinfo{edition}{second} edition.
\newblock \bibinfo{publisher}{Prentice Hall}.

\bibitemdeclare{book}{parteval}
\bibitem{parteval}
\bibinfo{author}{Neil~D. \surnamestart Jones\surnameend},
  \bibinfo{author}{Carsten~K. \surnamestart Gomard\surnameend} \&
  \bibinfo{author}{Peter \surnamestart Sestoft\surnameend}
  (\bibinfo{year}{1999}): \emph{\bibinfo{title}{Partial Evaluation and
  Automatic Program Generation}}.
\newblock \bibinfo{publisher}{Prentice Hall}.
\newblock \bibinfo{note}{\url{http://www.itu.dk/people/sestoft/pebook}}.

\bibitemdeclare{misc}{apt-www}
\bibitem{apt-www}
\bibinfo{author}{\surnamestart {Kestrel Institute}\surnameend}:
  \emph{\bibinfo{title}{{APT}}}.
\newblock \bibinfo{howpublished}{\url{https://www.kestrel.edu/research/apt}}.

\bibitemdeclare{misc}{specware-www}
\bibitem{specware-www}
\bibinfo{author}{\surnamestart {Kestrel Institute}\surnameend}:
  \emph{\bibinfo{title}{{Specware}}}.
\newblock \bibinfo{note}{\url{https://www.kestrel.edu/research/specware}}.

\bibitemdeclare{inproceedings}{java-light}
\bibitem{java-light}
\bibinfo{author}{Tobias \surnamestart Nipkow\surnameend} \&
  \bibinfo{author}{David \surnamestart von Oheimb\surnameend}
  (\bibinfo{year}{1998}): \emph{\bibinfo{title}{Java Light is Type-Safe ---
  Definitely}}.
\newblock In: {\sl \bibinfo{booktitle}{Proc.\ 25th {ACM} Symposium on
  Principles of Programming Languages ({POPL})}}, pp.
  \bibinfo{pages}{161--170}, \doi{10.1145/268946.268960}.

\bibitemdeclare{inproceedings}{pvs-codegen}
\bibitem{pvs-codegen}
\bibinfo{author}{Nararajan \surnamestart Shankar\surnameend}
  (\bibinfo{year}{2017}): \emph{\bibinfo{title}{A Brief Introduction to the
  {PVS2C} Code Generator}}.
\newblock In: {\sl \bibinfo{booktitle}{Proc.\ Workshop on Automated Formal
  Methods ({AFM'17})}}.

\bibitemdeclare{book}{zed}
\bibitem{zed}
\bibinfo{author}{J.~M. \surnamestart Spivey\surnameend} (\bibinfo{year}{1992}):
  \emph{\bibinfo{title}{The {Z} Notation: A Reference Manual}},
  \bibinfo{edition}{second} edition.
\newblock \bibinfo{publisher}{Prentice Hall}.

\bibitemdeclare{book}{jbook}
\bibitem{jbook}
\bibinfo{author}{Robert \surnamestart St{\"a}rk\surnameend},
  \bibinfo{author}{Joachim \surnamestart Schmid\surnameend} \&
  \bibinfo{author}{Egon \surnamestart B{\"o}rger\surnameend}
  (\bibinfo{year}{2001}): \emph{\bibinfo{title}{{Java} and the {Java} Virtual
  Machine: Definition, Verification, Validation}}.
\newblock \bibinfo{publisher}{Springer}, \doi{10.1007/978-3-642-59495-3}.

\bibitemdeclare{misc}{acl2-manual}
\bibitem{acl2-manual}
\bibinfo{author}{\surnamestart {The ACL2 Community}\surnameend}:
  \emph{\bibinfo{title}{The {ACL2} Theorem Prover and Community Books:
  Documentation}}.
\newblock \bibinfo{note}{\url{http://acl2.org/manual}}.

\bibitemdeclare{misc}{acl2-code}
\bibitem{acl2-code}
\bibinfo{author}{\surnamestart {The ACL2 Community}\surnameend}:
  \emph{\bibinfo{title}{The {ACL2} Theorem Prover and Community Books: Source
  Code}}.
\newblock \bibinfo{note}{\url{http://github.com/acl2/acl2}}.

\end{thebibliography}

%%%%%%%%%%%%%%%%%%%%%%%%%%%%%%%%%%%%%%%%%%%%%%%%%%%%%%%%%%%%%%%%%%%%%%%%%%%%%%%%

\end{document}